\documentstyle[epsf]{l-aa}
\newcommand{\gesim}{\,\raisebox{-0.4ex}{$\stackrel{>}{\scriptstyle\sim}$}\,}
\newcommand{\lesim}{\,\raisebox{-0.4ex}{$\stackrel{<}{\scriptstyle\sim}$}\,}
\newcommand{\rc}{r_{\rm c}}

\newcommand{\eacc}{E_{\rm acc}}
\newcommand{\eaccsd}{E_{\rm acc,Sd}}

\newcommand{\eloss}{E_{\rm loss}}

\newcommand{\ush}{u_{\rm sh}}
\newcommand{\uin}{u_{\rm sh,in}}
\newcommand{\tsd}{t_{\rm Sd}}

\begin{document}

   \thesaurus{06         
              (09.09.1 SN1006;  
               09.13.1;  
               13.07.3)} 
\title{TeV emission from SN 1006}

\author{A. Mastichiadis\inst{1} \and O.C. de Jager\inst{2}} 

\offprints{O.C. de Jager}

\institute{MPI Kernphysik, Postfach 10 39 80, D-69029 Heidelberg, Germany
      \and Space Research Unit, PU vir CHO, Potchefstroom, South Africa}

\date{Received ; accepted}

\maketitle

\begin{abstract}

Supernova 1006 is the first shell type supernova remnant
to show evidence of particle acceleration to TeV energies.
In the present paper we examine this possibility by modeling
the observed X-ray non-thermal emission in terms of synchrotron
radiation from Fermi accelerated electrons. The predicted synchrotron
spectrum fits the radio and non-thermal component of the
observed soft X-ray to hard X-ray emission quite well. 
These particles can produce TeV gamma rays by inverse Compton
scattering on the microwave radiation and other ambient fields, and 
the derived electron distribution is also used to calculate the
expected inverse Compton flux. We find that if the remnant
is characterised by a magnetic field strength lower than $\sim 7\mu$G, 
then the TeV flux can be higher than that of the Crab Nebula. 
About 75\% of the TeV emission from SN 1006 is expected to be concentrated 
in the synchrotron bright NE and SW rims (the ``hard aegis") of the remnant,
which would allow a sensitive search if the Atmospheric Imaging 
Cherenkov Technique is used. 

   \end{abstract}

\section{Introduction}

The type Ia supernova remnant G327.6+14.6, or SN 1006, is the remnant of
the explosion which took place in AD 1006. The distance to this
source is between 1.4 and 2.1 kpc (Green 1988). Its mean expansion rate
has been measured as $0.44\pm 0.13$ arcsec/yr, 
which implies a present expansion speed of 
$3700\pm 1300$ km/s for an average distance of 1.8 kpc. The 
time dependence of the 
expansion ($R\propto t^{0.48\pm 0.13}$) is consistent with Sedov
expansion, or with a forward/reverse shock pair moving into constant-density 
material (Moffett, Goss, \& Reynolds 1993).

The ASCA detection of power law X-ray 
emission from the bright northeastern (NE) 
and southwestern (SW) rims of SN 1006 led Koyama et al. (1995) to infer the
existence of electrons accelerated by the first order Fermi mechanism
up to energies of $\sim 200$ TeV. The detection of TeV $\gamma$-rays from
this system will prove the existence of such ultrarelativistic electrons
(De Jager et al. 1995, Mastichiadis 1996, Pohl 1996).

In Section 2 we will compile the non-thermal radio to X-ray spectrum, which
may be due to synchrotron emission from these electrons, and in
Section 3 we will compare the acceleration and loss timescale for
electrons accelerated by the first order Fermi process, and show that
the resulting maximum electron energy predicts a synchrotron cutoff
frequency which is consistent with observations. The best-fit spectrum
also leads to the derivation of the magnetic field 
strength $B_*$ associated with the shell.
In Section 4 we will calculate the inverse Compton (IC) $\gamma$-ray spectrum
associated with this best-fit model spectrum from Section 3. The
detection of TeV $\gamma$-rays from SN 1006 will enable the observer
to determine the magnetic field strength, and an important parameter 
associated with Fermi acceleration from the observed $\gamma$-ray flux.

\section{The observed synchrotron spectrum of SN 1006}

The radio morphology shows two bright
arcs towards the NE and SW (see e.g. Reynolds \& Gilmore 1993), with
a composite spectrum which is given by
\begin{equation}
\label{snu}
F_{\nu}=16(\frac{\nu}{10^9\;{\rm Hz}})^{-0.56}\;\;{\rm Jy},
\end{equation}
and shown in Fig. 1. The individual radio flux
measurements for the total emission from SN 1006
(reproduced from Fig. 1 of Reynolds 1996) are also shown for comparison.

ROSAT imaging
observations of SN 1006 shows that the soft X-ray remnant is dominated
by thermal ($kT=0.15$ keV) emission towards the SE, whereas the X-ray
counterparts of the NE and SW radio rims (or ``hard aegis") are characterized
by a power law spectrum which dominates the total emission from the remnant
above $\sim 1$ keV (Willingale et al. 1996). Spatially resolved spectral
results by ASCA have shown that these bright rims are responsible for
$\sim 75\%$ of the total X-ray flux from SN 1006 above 1 keV, and the interior
of the remnant also contains this non-thermal emission, which dominates
the thermal component above 2 keV (Koyama et al. 1995). The 
results from earlier non-imaging instruments are therefore useful to obtain
the non-thermal spectrum above 1 keV from the entire remnant. For example: 
TENMA (Koyama et al. 1987), EXOSAT (Jones \& Pye 1989), and Ginga
(Ozaki et al. 1994) obtained well-constrained spectral energy indices ranging
between $-$2.0 and $-$2.3 for energies between 1.5 keV and $\sim 10$ keV.
This is confirmed by the energy index of $-2.0\pm 0.2$ 
of the NE rim found by ASCA (Koyama et al. 1995).
The EXOSAT ME field of view was just large enough to cover the entire
remnant, without significant background contamination, and the energy
spectrum above 1.5 keV (derived from the given energy flux) is given by
\begin{equation}
\label{hard}
F_{\epsilon}=0.066\epsilon_{\rm keV}^{-2.0}\;\;{\rm keV.cm^{-2}s^{-1}keV^{-1}},
\end{equation}
which is shown together with the radio spectrum in Fig. 1.

The change from a $-$0.56 radio energy index to a $\sim -2$ X-ray energy index
above $\sim 1$ keV is indicative of a spectral turnover or break
around $\epsilon_b\sim 0.25$ keV, where the spectral energy index should be
around $-$1 (depending on the sharpness of the cutoff). In fact,
soft X-ray imaging observations of the hard aegis by the
EINSTEIN SSS (Becker et al. 1980) and EXOSAT LE (Jones \& Pye 1989) 
resulted in a consistent spectral energy index of $-$1.2. 
Willingale et al. (1996) used ROSAT PSPC observations to separate
the thermal and non-thermal energy fluxes of the hard aegis and the rest 
of the remnant, with the same finding as Koyama et al. (1995)
that the non-thermal soft X-ray flux from the
hard aegis contributes to $\sim 75\%$ of the total non-thermal soft X-ray 
flux of $2.66\times 10^{-10}$ ergs/cm$^2$/s (for the 0.1 keV to $\sim 2$ keV
range). Assuming the average spectral index of $-$1.2 found by 
EINSTEIN and EXOSAT
(Willingale et al. 1996 obtained a spectral index for the SW limb only),
but using the ROSAT normalization for the total non-thermal flux from
SN 1006, the resulting energy spectrum for this energy range is therefore
\begin{equation}
\label{soft}
F_{\epsilon}\sim 0.047\epsilon_{\rm keV}^{-1.2}
\;\;{\rm keV.cm^{-2}s^{-1}keV^{-1}},
\end{equation}
and is indicated by its spectral index in Fig. 1.

\section{First order Fermi acceleration of electrons in SN 1006}

The non-thermal synchrotron emission of SN1006 from shock accelerated
electrons has been discussed by Reynolds \& Chevalier (1981),
Ammosov et al. (1994) and, more recently, by Reynolds (1996). Here
we apply the method given in Mastichiadis (1996 - henceforth M96).
This allows us to treat the problem in a self-consistent manner
by calculating the electron distribution function at each instant in time 
from the solution of a time-dependent kinetic equation
for electrons. The physical picture 
is according to the `Onion-shell-model' (Bogdan \& V\"olk 1983)
as applied to time-dependent acceleration (Ball \& Kirk 1992).
Therefore we assume that electrons are accelerated in the 
(parallel) shock wave of the expanding
supernova remnant and at each instance there is a flux of relativistic electrons 
which escape downstream and subsequently radiate.
For these we assume to have a power law 
spectrum in energy,
i.e. $Q_{\rm e}(E,t)=Q_{\rm e,0}(t) E^{-s}$ with $E\le E_{\rm max}(t)$
and an exponential cut-off for $E > E_{\rm max}(t)$ (Webb et al. 1984).
For the determination of the normalization factor
we followed M96 (after an original suggestion by Drury 1992)
and set $Q_{\rm e,0}(t)\propto R(t)^2u_{\rm sh}(t)^3$
where $R(t)$ is the radius of the supernova and $\ush(t)$
is the velocity of the shock. 
Furthermore, the maximum energy electrons can achieve 
at a time t  depends on whether synchrotron losses become dominant or not
and it is given by the minimum of  
(Lagage \& Cessarsky 1983,
Webb et al. 1984, M96 and Reynolds 1996)
\begin{equation}
\eacc(t)\simeq 5.10^{-3}~A(\rc)f^{-1}u_{\rm sh}(t)^2Bt_{\rm yr}~{\rm erg} 
\label{eacc}
\end{equation}
and
\begin{equation}
\eloss (t)\simeq 2.10^{-4} f^{-1/2}A(\rc)^{1/2}\ush(t) B^{-1/2}~{\rm erg},
\label{elos}
\end{equation}
where $u_{\rm sh}$
is the shock velocity in units of km/sec, B is the magnetic field strength in Gauss,
$t_{\rm yr}$ is the supernova remnant age in years,
and $A(\rc)=(\rc-1)/\rc(\rc+1)$
with $\rc$ being the compression ratio of the shock connected with the
spectral index $s$ through the relation $s=(\rc+2)/(\rc-1)$. 
Also $f$ is the gyrofactor, i.e. the ratio of the particle's mean free path
to its gyroradius. The case $f=1$ corresponds to a simple Bohm diffusion.

The relevant energy losses which were included 
in the kinetic equation were adiabatic due to the 
expansion of the remnant, synchrotron and
inverse Compton. The only free parameter here is the magnetic field
as the photon fields on which the electrons lose energy by inverse Compton
scattering are rather well determined (see section 4 below).

Adopting standard supernova parameters, i.e. a total 
explosion energy of $W_{\rm SN}=
5.10^{50}W_{5,50}$ erg, an initial shock velocity of
$\uin=7.10^3u_{7,3}$ km/sec and an external matter density of 
$\rho=0.4\rho_{.4}$ H-atom/cm$^3$ (Willingale et al. 1996)
we solved numerically the 
electron kinetic equation which includes a source term and loss terms
as these were described above (for more details see M96).
The obtained electron 
distribution function was folded with the synchrotron emissivity 
to obtain the radiated spectrum as this should be observed at a
time $t_{\rm now}\simeq 1,000$ years. This way
we have made numerical fits to the radio/X-ray data by treating
as free parameters the strength of
the magnetic field downstream, the gyrofactor $f$
and the shock compression ratio $\rc$.

Fig.1 shows the best fit to the radio and X-ray spectra
(shown here as full lines) which was obtained 
for $B=3.5f^{2/3}~\mu$G and $\rc=3.73$ (corresponding to $s=2.1$).
This fit holds for $f\lesim 30$. As $f$ increases above
this value the spectrum starts to break due to synchrotron losses and
no satisfactory fit could be found. Therefore X-ray
data alone might not be able to constrain $f$ significantly. 
More constraints on $f$ can be placed 
from future TeV observations as we will show in the next section.

We proceed now by giving a simple qualitative picture 
which will help
in the understanding of our results: According to the standard
picture of cosmic ray acceleration in supernova remnants (see, for example, 
Dorfi 1991)
both $Q_{\rm {e,0}}$ and $\eacc$ peak close to the transition 
of the supernova remnant from its free expansion to its 
Sedov phase which occurs at 
$\tsd\simeq 410 W_{5,50}^{1/3}\rho_{0.4}^{-1/3}
u_{7,3}^{-5/3}~~{\rm yr}$.
For example, $\eacc\propto t$ for $t<\tsd$ and 
$\eacc\propto t^{-1/5}$ for $t>\tsd$. 
Similarly, $Q_{\rm {e,0}}\propto t^2$ for $t< \tsd$ and $Q_{\rm {e,0}}\propto t^{-1}$
for $t> \tsd$.
Therefore the radiation observed now should be dominated
by the particles accelerated at $t\gesim\tsd$.
Using Eqn. (4) we get that the maximum particle energy at $t\simeq\tsd$
is given by
\begin{equation}
\eaccsd\simeq 1.5~10^7 Bf^{-1}g^{-1}W_{5,50}^{1/3}
\rho_{0.4}^{-1/3}u_{7,3}^{1/3}~{\rm erg}.
\label{esedov}
\end{equation}
$g$ is a factor (of order unity) that reflects the fact that
at $t=\tsd$ the velocity is not any more $\uin$ but somewhat
smaller since deceleration of the shock should have started before that time. This 
in turn implies that the energy $\eacc$ as given by Eqn. (\ref{esedov})  
is, strictly 
speaking, an upper limit of the maximum energy the electrons can
attain during acceleration.  

Equation (\ref{esedov}) is the maximum energy electrons can attain
as long as this is smaller than $\eloss$ (equation [\ref{elos}]).
A comparison of the two expressions for $t=\tsd$
yields that this is indeed the case
as long as $B\lesim 10^{-5}f^{1/3}g^{1/3}W_{5,50}^{2/9}\rho_{0.4}^{-2/9}
u_{3,7}^{4/9}$~G.
If this relation is satisfied then synchrotron losses do not become
important at {\sl any} epoch, since for $t>\tsd$, $\eacc\propto t^{-1/5}$
while $\eloss\propto t^{-3/5}$.

>From the above discussion it becomes evident that if radiation
losses do not become important then the synchrotron spectrum
that the current model predicts
can be fairly well represented by the 
synchrotron spectrum which is derived from 
a simple electron distribution function (power law times exponential cutoff)
of the form 
\begin{equation}
\label{best}
\frac{dN}{dE}\simeq 1.5\times 10^3 (\frac{B}{B_*})^{-1.56}E^{-2.12}
\exp{\left[-\frac{E}{E_b}
(\frac{B}{B_*})^{1/2}\right]}
4\pi d^2,
\end{equation}
with units in total number of electrons per erg while $E$ is in ergs.
The best-fit value for the magnetic field is given by
$B_*=3.5f^{2/3}g^{2/3}W_{5,50}^{-2/9}\rho_{0.4}^{-2/9}u_{7,3}^{2/3}$ 
$\mu$G while 
$E_b=\eaccsd$. The area factor $4\pi d^2$ 
allows us to calculate photon spectra in terms of the flux at Earth if
$(dN/dE)/4\pi d^2$ is used. 
Using the best fit value for $B_*$ 
the total energy content in relativistic electrons is
$\simeq 10^{49}~f^{-1.04}g^{-1.04}W_{5,50}^{0.35}\rho_{0.4}^{-0.35}u_{7,3}^
{0.35}$ erg. This in turn gives us an 
efficiency of cosmic ray electron production of
\begin{equation}
\eta_{\rm el}\simeq 0.02f^{-1.04}g^{-1.04}W_{5,50}^{-0.65}\rho_{0.4}^{-0.35}
u_{7,3}^{0.35}
\end{equation}
in rough agreement with the hypothesis that about a few percent
of the supernova energy available for acceleration goes to 
electrons.

\begin{figure}[t]
\vspace{2 cm}
\epsfxsize=8.5 cm
\epsffile{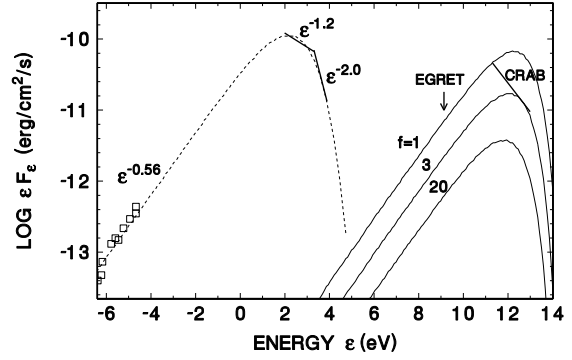}
\vspace{-5 cm}
\caption{Plot of the best fit synchrotron spectrum (thin dashed line) 
and inverse Compton spectra (thin solid lines indicated by the $f$-values
as described in Section 3) calculated from the time-dependent electron 
continuity equation (Section 3). The open squares are the radio 
flux measurements reproduced from Reynolds (1996).
The thick solid lines represent the observed X-ray spectra 
corresponding to Eqns. (\protect\ref{hard}), and (\protect\ref{soft})
as discussed in Section 2, and indicated by their spectral indices.
The thick dashed line is the Crab spectrum given by Djannati-Ata\"i (1995).}
\end{figure}

\section{The expected $\gamma$-ray spectrum of SN 1006}
Several soft photon fields may contribute to the 
inverse Compton (IC) scattering of relativistic electrons into the
$\gamma$-ray region. De Jager (1996) discussed the production of
$\gamma$-ray spectral features introduced by the scattering of
several thermal photon spectra by the electron cutoff (with maximum electron
energy $\gamma_bmc^2$) in SNR W44,
which may explain the EGRET detection (2EG J1857+0118, Thompson et al. 1996)
of this remnant. The cutoff in the spectrum derived in Section 3 
will similarly produce $\gamma$-ray bumps on a 
$\epsilon F_{\epsilon}$ plot (as in Fig. 1) at $\gamma$-ray energies
$\sim 3kT_i\gamma_b^2$ (in the Thomson limit) or $\sim \gamma_b mc^2$
(in the Klein-Nishina limit), with $T_1=2.76$K representing the
CMBR, $T_2\sim 25$K the galactic dust contribution, $T_3=3,000-4,000$K
representing the Population II stars, and $T_4\sim 7,500$K
representing the contribution from Population I stars, with
densities given by Skibo (1993). Arendt (1989) also did not detect
any significant FIR emission associated with swept up dust in SN
1006, so that there would be no contribution to soft photons from
sources inside or near SN 1006, in contrast with W44 where the 
emission from swept-up dust dominates all soft photon fields.

We include all soft photon fields,
using the full cross section for isotropic IC
as reviewed by Blumenthal \& Gould (1970) in the calculation of the
$\gamma$-ray spectrum.
The  IC spectra calculated from the best-fit electron spectrum 
(Eqn.~[\ref{best}]) for different $f$-values are also
shown in Fig. 1, and it is clear that the $\gamma$-ray energy flux peaks 
at a few TeV (nearly independent of the value of $f$), the dominant
source of this bump being the CMBR photons scattered in the Thomson limit.
The integral flux above 1 TeV is given by 
\begin{equation}
\label{ftev}
F(>1{\rm TeV})=4.1\times 10^{-11}f^{-1.35}\;\;{\rm cm^{-2}s^{-1}},
\end{equation}
with $f$ constrained  to values below $\sim 20$.
The Crab spectrum in the range 0.5 to 10 TeV 
(Djannati-Ata\"i 1995) is also shown for comparison
in Fig. 1, and it is clear that we may expect SN 1006 to be a stronger 
source than the Crab at 1 TeV if $f<3$.
It is also clear that the EGRET upper limit of SN 1006 shown in Fig. 1
does not constrain $f$ significantly, but TeV observations will provide
valuable limits.

Finally we would like to note that the TeV 
flux as given above
is about an order of magnitude higher than the expected
flux from the nuclear component of cosmic rays assumed
to be accelerated also at the supernova shock (Drury et al. 1994).

\section{Summary-Discussion}

The recent X-ray observations of SN 1006 (Koyama et al. 1995) make a 
strong case for particle acceleration in supernova remnants.
A necessary consequence of this is that SN 1006 can be a 
source of TeV radiation as the accelerated electrons will scatter
off the ambient soft photons to very high energies.
In the present paper we calculated the expected TeV flux by 
following the basic principles
of first order Fermi shock acceleration to fit the radio/X-ray
observations and consequently we used the derived electron
distribution function to calculate the TeV emissivity. 
In this case we found that the TeV flux depends sensitively on
the gyrofactor $f$ of the accelerated electrons and consequently one
can use any future TeV observations of SN 1006 to put limits
on this. 

We find that the EGRET upper limit (derived from Thompson et al. 1996)
does not constrain $f$ significantly, but imaging TeV observations
have no difficulties observing sources nearly 10 times weaker than the
Crab. We may therefore be able to detect SN 1006 at TeV energies if $f$
is not too large. The NE and SW limbs may be detectable by the Atmospheric
Imaging Cherenkov Technique (see e.g. Weekes et al. 1989) if a search for
the NE and SW rims are made. However, 
the flux for each rim  would be $\sim 38\%$ of the flux
given in Eqn. (\ref{ftev}) for the total TeV emission from SN 1006, but
by superimposing the pixels from the TeV image
corresponding to the NE and SW rims, the total flux should be about
75\% of the flux given in Eqn. (\ref{ftev}).

\acknowledgements
AM would like to thank 
the Deutsche Forschungsgemeinschaft for support under
Sonderforschungsbereich 328 and the staff and members 
of the Space Research Unit of the
Potchefstroom University for their hospitality.


\end{document}